\documentclass[useAMS,usenatbib]{mn2e}
\usepackage{graphicx,times}

\title[]{The missing compact star of SN1987A: a solid quark star?}
\author[]{X. W. Liu$^{1}$\thanks{E-mail:
xiongwliu@163.com}, J. D. Liang$^{2}$, R. X. Xu$^{1}$, J. L. Han$^{3}$, and G. J. Qiao$^{1}$\\
$^{1}$School of Physics and State Key Laboratory of Nuclear Physics and Technology,
              Peking University, Beijing 100871, China\\
$^{2}$Center for Astrophysics, Tsinghua University, Beijing 100084, China\\
$^{3}$National Astronomical Observatories, Chinese Academy of Sciences, Beijing 100012, China}
\begin{document}

\date{Accepted. Received}

\pagerange{\pageref{firstpage}--\pageref{lastpage}} \pubyear{2012}

\maketitle

\label{firstpage}

\begin{abstract}
   To investigate the missing compact star of Supernova 1987A, we analyzed both the cooling and the heating processes of a
   possible compact star based on the upper limit of observational X-ray luminosity. From the cooling process we found that a solid
   quark-cluster star, which has a stiffer equation of state than that of conventional liquid quark star, has a heat capacity
   much smaller than a neutron star. It can cool down quickly, which can naturally
   explain the non-detection of a point source (neutron star or quark star) in X-ray band. On the other hand, we consider
   the heating process from magnetospheric activity and
   possible accretion, and obtain some constraints to the parameters of a possible pulsar. We conclude that
   a solid quark-cluster star can be fine with the observational limit in a large and acceptable parameter space.
   A pulsar with a short period and a strong magnetic field (or with a long period and a weak field) would has luminosity higher than the luminosity limit if the optical depth is not large enough to hide the compact star. The constraints of the pulsar parameters can be tested if the central compact object in 1987A is discovered by advanced facilities in the future.
\end{abstract}


\section{Introduction}

   Pulsars are thought to be neutron stars (NSs) since the discovery of the first
   pulsar \citep{Hewish}. Nevertheless, the equation of state (EoS) of neutron stars is not clear till now.
   \citet{Witten} pointed out that the true
   ground state of hadrons may be strange matter, which contains roughly equal numbers of up, down and strange
   quarks. It is then realized that a quark star or strange star (SS) might be the ground state
   of NSs \citep{Alcock,Haensel}. NSs may convert to SSs.

   Are pulsars NSs or SSs? Essentially,
   this is a problem of non-perturbative quantum chromodynamics (QCD), which describes the strong interaction processes
   in low energy scales, and is very hard to solve in mathematics. Fortunately, the astronomical
   observations could discriminate NSs and SSs, and help to understand the non-perturbative QCD issue.

   There are many methods to constrain the EoSs of pulsars and check the differences between NSs and SSs.
   For example, SSs are self-constrainted, while NSs are gravity constrainted, so that they have
   different mass-radius relations. If we can measure the mass and radius of a pulsar, we can discriminate whether it is a NS or a SS.
   This has not succeeded up to now, mainly because of uncertainties of the radii.
   Nevertheless, \citet{Li} suggested that SAX J1808.4-3658 may be a quark star based on this relation and available data.
   Another constraint is the minimum mass of a pulsar or
   sub-millisecond pulsar. Since SSs are self-bounded, their mass can be
   much smaller than the minimum mass of NS, which could be about $0.1M_{\odot}$ and can spin with a period shorter than 1 millisecond
   while a NS can not \citep{Du}.
   If the mass of a pulsar is smaller than 0.1$M_{\odot}$ or the period is shorter than 1 millisecond,
   it could not be a NS.

   Different EoS of a pulsars gives a different
   maximum mass. The newly confirmed mass of PSR J1614-2230 of 2$M_{\odot}$ \citep{Demorest}
   rules out almost all currently proposed hyperon or
   boson condensate EoSs \citep{Lattimer}
   and traditional soft EoS of SS \citep{Chan}.
   \citet{Lai09} argued that the
   solid quark-cluster stars (SQS), a special kind of SS, could have
   the maximum mass to be larger than 2 $M_{\odot}$ because of the very stiff EoS.

   The bolometric radiation of a young pulsar can be used to
   distinguish NSs and SSs, because they are different in thermal capacity and surface radiation.
   We note that a SQS has very small thermal capacity because its temperature is much lower than the
   Debye temperature \citep{Yu}, it can cool down very fast and has a very lower bolometric
   luminosity than a NS. The missing compact object in SN1987A \citep{McCray} is studied in this paper with this luminosity constraint. 

   There is no doubt that SN1987A provides an unprecedented
   opportunity in astronomy and astrophysics studies. Yet the possible
   compact star produced in this explosion is still a mystery. On 23 February 1987,
   both the Kamiokande II detector and the Irvine-Michigan-Brookhaven
   detector observed a neutrino burst \citep{Hirata,Bionta} from the Large Magellanic Cloud,
   just before the optical shine. The energy spectrum
   and the flux density of the neutrino burst are consistent with the current
   theory of core-collapse supernova with an energy release of $\sim$3$\times$10$^{53}$ ergs,
   from which a NS is expected to form therein. Astronomers were excited
   by the possibility of watching the NS birth at the very beginning. From then on,
   the powerful detectors or telescopes on the ground and in the space searched it from radio
   to gamma ray bands, one after another. Unfortunately, no pulses were detected
   or no point source is found there \citep{McCray,Manchester}.

   Parkes 64-m radio telescope was used to search pulses from SN1987A in the first few years at frequencies between 400 and 5000 MHz,
   down to the limit of $\sim$0.2 mJy at 1.5 GHz \citep{Manchester88}.
   The strong efforts were made with Parkes in 2006 in several bands, but
   no pulsar candidates with a S/N ratio greater than 9.0 were found \citep{Manchester}.
   Some optical observations announced the detection of a pulsar in SN1987A \citep{Kristian, Murdin, Middleditch}, but all of them
   have not been confirmed. \citet{Percival} made an optical search using the Hubble Space Telescope.
   No significant pulsations were found in the period range of 0.2 ms to 10 s with an upper limit
   for the pulsed emission equivalent to a V magnitude of $\sim$24.
   A similar search was made by \citet{Manchester96} using the 3.9-m Anglo-Australian Telescope with similar parameters, and they got null results.
   \citet{Shtykovskiy} obtained a luminosity limit in the 2-10 keV band of $5\times10^{34}$ erg s$^{-1}$ using the XMM-Newton.
   \citet{Park02,Par} obtained upper limits of $5.5\times10^{33}$ erg s$^{-1}$ and $1.5\times10^{34}$ erg s$^{-1}$ in the same X-ray band using Chandra.

   To explain the non-detection of the predicted NS, several possibilities were discussed. First, some of the ejected material of SN may fall back to the
   NS surface shortly after the supernova explosion. We can not see the NS since it has been converted to a BH due to the fall back material.
   But this possibility should be small.
   The discovery of a 2 $M_{\odot}$ pulsar \citep{Demorest} suggests the maximum mass of pulsar could be much larger than 1.4 $M_{\odot}$, and therefore
   a normal NS needs to accrete a lot of material for the conversion to a BH.
   Another possibility is that the NS may be located in the cold dust cloud \citep {McCray}
   at the center of SN1987A, which may be opaque in some bands. But if
   the NS is not inside the dust or the dust is optical thin in X-ray band, it could be very
   intriguing because the observed upper limit at around 1 keV is already lower than
   the expected luminosity of a cooling NS with no heating \citep{Park,McCray}. \citet{Chan}
   suggested that the compact star at the center of SN1987A may be not a NS but a SS
   with a softer EoS than that of neutron star matter, which
   could have a X-ray luminosity less than 10$^{34}$ erg s$^{-1}$ at the age of 20 years.
   Note that the SS of this EoS is almost ruled out by the newly found 2 $M_{\odot}$ NS \citep{Demorest}, because its
   maximum mass is smaller than 2 $M_{\odot}$ due to its soft EoS.

   We emphasize that a model for quark star with a stiff EoS, e.g. the SQS \citep{Lai}, is still survived theoretically.
   The SQS is suggested by \citet{Xu} and improved by \citet{Zhou}, \citet{Lai09} and \citet{Lai}. The key point is that the interaction energy of
   quarks in a compact star could be higher than the Fermi energy when the density is lower than a few tens of nuclear densities \citep{Xu11}, so that
   quarks may be clustered and the star could be a solid star, i.e. the solid quark star (or quark-cluster star). The SQS model can naturally
   explain most of the observational features of pulsars \citep{Xu11}.
   The surface of SQS is self-bounded which provides a larger binding energy than the gravity-bounded NS. That is more helpful to the
   generation of sparks for radio emission \citep{Xu99,Qiao}. The X-ray emission and pulsations in
   magnetosphere are similar to those of a NS. The star quake of SQS can induce glitches and energetic bursts which may be shown as soft gamma-ray repeaters
   \citep{Xu06}. The SQS could has a larger maximum mass than other model of quark stars.
   \citet{Lai09} gives a maximum mass $>2 M_\odot$, which stands the test of the 2 $M_\odot$ pulsar \citep{Demorest}. The SQS has a smaller radius than any NSs when the mass is small. The newly discovery of the radius and mass of the rapid burster (MXB 1730-335) \citep{Sala} could naturally fit the SQS model.
   Additionally, the SQS has a very low heat capacity \citep{Yu} which helps to cool down very fast. This may explain the non-detection of
   a point source in SN1987A.
  
   In this work we will first analyze the cooling process of a SQS in comparison with
   NS and traditional liquid SS, and then further study the constraints on the parameters via heating processes of pulsars.
   Our conclusions and discussions are presented in the last section.

\section[]{Cooling of the possible compact object}
   A stellar BH has no classical radiation, and its Hawking radiation is
   negligible in astronomy, thus its cooling luminosity is nearly zero. For NSs and
   SSs, the cooling processes are determined by their heat capacities and surface radiations.

   The heat capacities of NSs, conventional SSs and SQSs are \citep{Maxwell,Ng,Yu},
   \begin{equation}
      C_{\mathrm{NS}} = C_{\mathrm{NS}}^{\mathrm{n}}+C_{\mathrm{NS}}^{\mathrm{e}},
   \end{equation}

   \begin{equation}
      C_{\mathrm{SS}} = C_{\mathrm{SS}}^{\mathrm{q}}+C_{\mathrm{SS}}^{\mathrm{g-\gamma}}+C_{\mathrm{SS}}^{\mathrm{e}},
   \end{equation}

   \begin{equation}
      C_{\mathrm{SQS}} = C_{\mathrm{SQS}}^{\mathrm{l}}+C_{\mathrm{SQS}}^{\mathrm{e}},
   \end{equation}
   where the superscripts n, e, q, g-$\gamma$, and l
   denote the contributions of neutrons, electrons, quarks, quark-gluon plasma and lattice structure, respectively.
   In eqs. (1) and (2), $C_{\mathrm{NS}}^{\mathrm{n}}$ and $C_{\mathrm{SS}}^{\mathrm{q}}$ are larger than
   $C_{\mathrm{NS}}^{\mathrm{e}}$ and $C_{\mathrm{SS}}^{\mathrm{e}}$ when temperature
   is higher than critical temperature $T_c$ ($\sim$10$^{9}$ K). When $T<T_c$ the superfluid state appears,
   $C_{\mathrm{NS}}^{\mathrm{n}}$ and $C_{\mathrm{SS}}^{\mathrm{q}}$ will
   exponential decay and vanish quickly. In eqs. (2) and (3), both $C_{\mathrm{SS}}^{\mathrm{g-\gamma}}$ and
   $C_{\mathrm{SQS}}^{\mathrm{l}}$ are in proportion to $T^3$, while $C_{\mathrm{SS}}^{\mathrm{e}}$ and
   $C_{\mathrm{SQS}}^{\mathrm{e}}$ are in proportion to $T$. Thus the heat capacity of compact stars is
   dominated by electrons when temperature is not too high, with $T_c\sim$10$^{9}$K for NS and SS,
   and $\sim$10$^{10}$K for SQS \citep{Yu}. The temperature will otherwise quickly cool down below $10^9$K
   within dozens of seconds. Therefore, the heat capacity of electrons is overwhelming the cooling process in almost all observational time.
   The heat capacity of electrons in NS, SS and SQS are \citep{Maxwell,Ng,Yu},
   \begin{equation}
      C_{\mathrm{NS}}^{\mathrm{e}}=1.9\times10^{37}M_1\rho_{14}^{1/3}T_9 \mathrm{~erg~K^{-1}},
   \end{equation}

   \begin{equation}
      c_{\mathrm{SS}}^{\mathrm{e}}=1.7\times10^{20}(Y\rho/\rho_0)^{2/3}T_9 \mathrm{~erg~(cm^3~K)^{-1}},
   \end{equation}

   \begin{equation}
      C_{\mathrm{SQS}}^{\mathrm{e}} \simeq N_e \frac{k_B T_s}{E_F}\ k_B,
   \end{equation}
   where $M_1=M/M_\odot$, $\rho_{14}=\rho/10^{14}$ g cm$^{-3}$, $T_9=T/10^9$ K, $Y$ is the electron fraction, $\rho_0$ is nuclear matter density,
   $N_e$ is the electron number in a star, $k_B$ is Boltzmann's constant, $T_s$ is the value in the
   star's local reference frame, and $E_F$ is the Fermi energy of the degenerate electron gas.
   In the extremely relativistic case, $E_F=(\frac{3n_eh^3}{8\pi})^{1/3}c$, where $n_e$ is the number density of electrons,
   $h$ is the Planck constant, and $c$ is the speed of light. From eq. (6) we have
   \begin{equation}
      C_{\mathrm{SQS}}^{\mathrm{e}} \simeq 3.5\times10^{37}(YM_1)^{2/3}R_6T_9  \mathrm{~erg~K^{-1}},
   \end{equation}
   where $R_6$ is the star radius in units of $10^6$ cm. The electron number in quark star is much smaller than that in neutron star,
   $Y\sim 10^{-5}$. Thus the heat energy conserved in a quark star is far less than in a neutron star.

   Surface radiation is important in the cooling process. On the NS surface,
   the radiation can be simply treated as approximate black body radiation because of atomic atmosphere. On SS or SQS
   surface there is no atomic atmosphere, therefore the radiation mainly depends on the interaction between electric layer
   and photons. \citet{Chan} suggested that the surface radiation of SS is via bremsstrahlung, and predicted a luminosity smaller than
   $10^{34}$ erg s$^{-1}$ when it is older than 20 years.
   In their paper the bremsstrahlung calculations of \citet{Jaukumar} were used. We noticed that \citet{Caron} got a much low flux of the SS
   bremsstrahlung emission. In our work we adopt the formula of \citet{Caron} to calculate the cooling of SQS, which makes our results more reliable.
   Figure 1 shows the bremsstrahlung cooling curves of SQSs together with those of black body radiation of the model of \citet{La}. We
   ignored neutrino radiation and color superconductivity related photon emission mechanisms. It shows that the cooling luminosity of a SQS
   could be smaller than $10^{34}$ ergs s$^{-1}$ about 20 years after its birth even it cools down by bremsstrahlung emission.
   The SQS was not detected previously, when it has a larger luminosity, because of the deep X-ray optical depth \citep{McCray} and poor observation devices.
   Thus we can conclude that if there is a cooling SQS in SN1987A it should be undetectable up to now, which is compatible with the observations.
  
   \begin{figure}
   \centering
   \includegraphics[width=9cm]{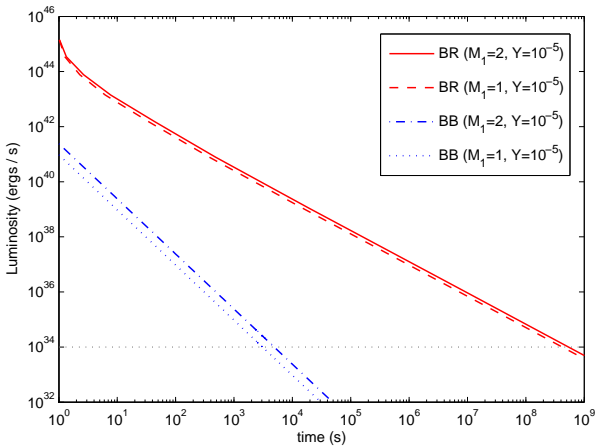}
      \caption{
      The cooling curves of SQSs with bremsstrahlung (BR) \citep{Caron} and black body radiation (BB). Neutrino radiation and color superconductivity
      related photon emission are not considered. The observational upper limit is indicated as a horizontal dotted line.
      It shows that the cooling luminosity of a SQS could be smaller than $10^{34}$ ergs s$^{-1}$ about 20 years after its birth even it
      cools down by bremsstrahlung emission.
      Here we take the stellar mass $M = M_1 M_\odot$
      and the number ratio of electron to baryon as $Y$.}
         \label{FigVibStab}
   \end{figure}

\section{Constraints on the parameters via heating processes}

   The bolometric luminosity of a compact star comes not only the contribution of cooling process but also heating
   processes from the activity of magnetosphere and the accretion from interstellar medium and accretion disks.
   The heating luminosity is probably lower than the upper limit of observations.
   The heating mechanisms are independent of
   the EoS of pulsar, therefore we can make some constraints on the physical parameters of a possible pulsar regardless it is a NS or a SQS.

   Usually, the activity of magnetosphere can produce both thermal and non-thermal X-ray emissions. When particles are
   accelerated in the magnetosphere they can emit non-thermal X-rays with a power law spectrum. When particles flow in and bombard on
   the pulsar surface, they will heat it, and thermal X-rays come out from the heated area.
   \citet{Becker} found that the non-thermal X-ray luminosity $L_x$ and spin down energy loss rate $\dot E$ are roughly related by
   $L_x=10^{-3}\dot{E}$. \citet{Yu} found that the thermal X-ray luminosity has a similar relation with that $\dot{E}$,
   $L_{bol}^{\infty}\sim10^{-3}\dot{E}$. Since the observations have given a upper limit of thermal X-ray luminosity
   $L_{bol}^{\infty}\sim10^{34}$ erg s$^{-1}$ \citep{Park02,Par},
   we can get an upper limit of spin energy loss rate $\dot{E}$. And we can further make a constraint to the parameters of
   spin and magnetic field of the possible pulsar, because $\dot{E}$ is the function of spin and magnetic field,
   \begin{equation}
   \dot{E}=-\frac{2} {3c^3}\ \mu_\perp^2\Omega^4,
   \end{equation}
   where $\mu_\perp$ and $\Omega$ are the vertical fraction of magnetic moment $\mu$ and the spin
   angular frequency of the pulsar. We use a parameter $a$ for the factor of thermal X-ray luminosity in terms of spin down energy lost rate, i.e.
   \begin{equation}
   L_{bol}^{\infty}=a\dot{E}.
   \end{equation}
   And then we get
   \begin{equation}
   \mu \simeq 3.22 \times 10^{14}a^{-1/2}{L_{bol}^{\infty}}^{1/2} P^{2},
   \end{equation}
   where $P=2\pi /\Omega$ is the spin period of a pulsar.

   Note that most of pulsars have an X-ray luminosity in the region of $10^{-4}-10^{-2}\dot{E}$ in both thermal and nonthermal cases
   \citep{Becker,Yu}. We used $a=10^{-4}$, $10^{-3}$ and $10^{-2}$ to get the upper left three lines in two panels of Figure 2. In the figure,
   the region to the left of the dash-dotted line should
   be ruled out because even the X-ray factor is as small as $10^{-4}$, it's luminosity should be larger than $10^{34}$ erg s$^{-1}$.
   The region below the solid line should be compatible with observational upper limit, because the X-ray luminosity is smaller than $10^{34}$ erg s$^{-1}$
   even $a=10^{-2}$.

   The accretion heating can also be used to restrict pulsar parameters. The accretion from interstellar
   medium is intensively dependent on the proper motion of a pulsar, and usually has a very small luminosity which we can ignore. Here we consider only
   the possible accretion from the fall back disk.

   To understand the evolution of the fall back disk we define three radius: the light speed radii $r_L$, the co-rotation radius
   $r_{co}$ and the magnetosphere radius $r_m$, as,
   \begin{equation}
    r_L=cP/2\pi,
   \end{equation}
   \begin{equation}
    r_{co} =(GM/4\pi^2)^{1/3}P^{2/3},
   \end{equation}
   \begin{equation}
   r_m = \mu^{4/7}(2GM)^{-1/7} {\dot M}^{-2/7},
   \end{equation}
   where $G$, $M$ and $\dot M$ are the gravitational constant, the mass of pulsar and the accretion rate, respectively.
   In fall back disk regime a pulsar often undergoes three phases: pulse phase, propeller phase and accretion phase.
   The pulse phase happens in the early stage of a pulsar, when a young pulsar has a fast spin and strong radiation which
   pushes the interstellar medium out of the light speed radius $r_L$, thus the compact star acts as a pulsar. When the radiation
   becomes not so strong, some of the medium may go into the light radius and interact with the magnetosphere. When $r_{co}<r_m<r_L$ or even
   $r_m$ is a little smaller than $r_{co}$, the magnetic freezing effect would force the flow in medium to co-rotate with the pulsar,
   and the spin energy of a pulsar will transfer to the medium via the interaction. This is so called the propeller phase.
   In this phase only small amount of medium can diffuse and fall on the star surface, which induces the low luminosity X-ray emission.
   If the accretion material can across the co-rotation radius and access the so called break radius $r_{br}$, which is a little smaller
   than $r_{co}$ (i.e. $r_m<r_{br}<r_{co}$), a massive accretion would occur. This is the accretion phase, which usually produces
   a very large X-ray luminosity.

   The Hubble Space Telescope found dust clouds interior to the debris in SN1987A \citep {McCray}, which could be
   indication of the formation of a fall back disk around the centra compact star. If the fall back disk does exist, we would expect
   the pulsar is not in accretion phase which otherwise results in X-ray luminosity larger than $10^{34}$ erg s$^{-1}$.
   Thus the disk should have $r_m>r_{br}$. We presume $r_{br}$ is proportion to $r_{co}$, i.e. $r_{br}=br_{co}$,
   here $b$ is a constant and $0<b<1$.
   Therefore to fit with the observations we need
   \begin{equation}
      r_m > br_{co}.
   \end{equation}
   Submitting $r_m$ and $r_{co}$ to the above inequation, we get the relation of the pulsar parameters,
   \begin{equation}
      \mu_{30} > 0.074b^{7/4}{M_1}^{5/6}{\dot M}_{16}^{1/2}P^{7/6},
   \end{equation}
   where $\mu_{30}=\mu /10^{30}$ and ${\dot M}_{16}=\dot M/10^{16}$.

   \begin{figure}
   \centering
   \includegraphics[width=9cm]{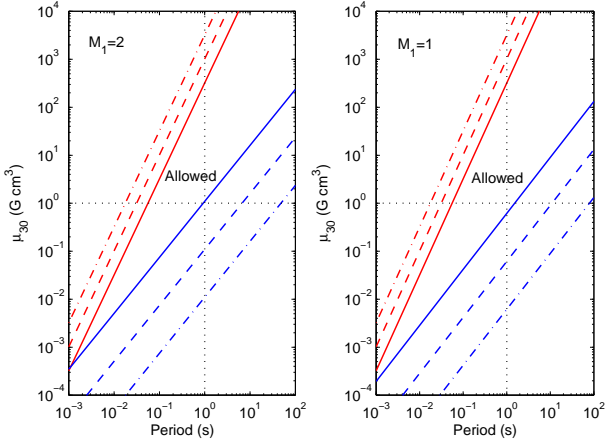}
      \caption{Constraints on parameters of the possible compact object in SN1987A via magnetosphere action heating (upper left three lines)
      and fall back disk accretion heating (lower right three lines). Left panel: $M_1=2$ and $b=0.9$, $a=10^{-4}$
      for red dash-dotted line, $a=10^{-3}$ for red dashed line, $a=10^{-2}$ for red solid line, $\dot{M}=10^{18}$ erg s$^{-1}$ for blue
      solid line, $\dot{M}=10^{16}$ erg s$^{-1}$ for blue dashed line and $\dot{M}=10^{14}$ erg s$^{-1}$ for blue dash-dotted line. Right panel:
      Same as the left but $M_1=1$.}
         \label{FigVibStab}
   \end{figure}

   The lower right three lines in both panels of Figure 2 show the magnetic moment depend on the spin period with other given parameters.
   We presumed $b=0.9$ in both cases, $M_1=2$ for left and $M_1=1$ for right. We used the accretion rate from ${\dot M}=10^{18}$ g s$^{-1}$ to
   ${\dot M}=10^{14}$ g s$^{-1}$. It should be noticed that when ${\dot M}<10^{14}$ g s$^{-1}$ the heating
   luminosity can not exceed $10^{34}$ erg s$^{-1}$
   even in the accretion phase. For a disk around a pulsar the accretion rate is almost impossible to exceed $10^{18}$ erg s$^{-1}$.

   Therefore, two kinds of heating mechanisms provide two sets of limits to the parameters
   of a possible pulsar, as shown in Figure 2. From the combination of both heating mechanisms, the region between two solid lines is fine with observations,
   while the upper left and lower right regions are the forbidden areas. From the comparison of left panel and right panel
   we can see that pulsars with a smaller mass has a wider region of parameters than the massive one.
   Because the heating mechanisms are EoS independent, our results are applicable to both NS and SQS.
\section{Conclusions AND DISCUSSIONS}

   We analyzed the cooling processes of compact stars and found that the SQSs could have a thermal luminosity lower than the
   observational upper limit of SN1987A. We studied the possible heating processes of a young pulsar and obtained some
   constraints on the pulsar parameters. It should be reasonable to conclude that:

   \begin{enumerate}
      \item A solid quark-cluster star with normal parameters is compatible with the non-detection in SN1987A,
      because both its cooling and heating luminosities should be lower than the observational upper limit. A low mass quark star has a wider
      parameter space than a massive one.
      \item If the compact star is shielded by dust, the parameter constraints should be relaxed. The parameter space of a SQS can be wider, even the object could be a normal NS.
      \item However, a black hole candidate can not be ruled out by cooling and heating analysis,
      though it is very difficult to form via accretion as we mentioned before.
    \end{enumerate}

   The predicted parameter space in this paper could be tested by the future observations.
   The impact of the supernova blast wave with its circumstellar matter is producing a ring which is visible
   from mm-band to X-ray band of SN1987A \citep{Bouchet,Gaensler,McCray,France} and it is still brightening. This ring would
   make it almost impossible to detect or rule out
   a cooling compact star in these bands. But a heating pulsar may be detected in the future because the heating luminosity from
   magnetospheric activity decays very slowly, and when the fall back disk evolves into the accretion phase so that the X-ray luminosity could be
   much larger than that in other phases and finally it could be detectable. It is not clear whether
   the low frequency radio and high energy $\gamma$-ray bands are affected by the ring which preserves the possibility of discovering
   the compact star and test the parameters by the future advanced facilities, e.g. the Square Kilometer Array (SKA) telescope.

\section*{Acknowledgments}

We would like to thank the referee for helpful comments and the pulsar group of PKU for useful discussions.
This work is supported by the National Natural Science
Foundation of China (Grant Nos. 10935001, 10973002, 10833003), the National
Basic Research Program of China (Grant Nos. 2009CB824800,
2012CB821800), and the John Templeton Foundation.

\end{document}